\documentclass{scrartcl}
\usepackage{amsmath}
\usepackage{graphicx}
\usepackage{cite}
\usepackage{amsfonts}
\usepackage{amssymb}

\def\sfrac#1#2{{\textstyle{#1\over #2}}}
\newcommand{\be}{\begin{equation}}
\newcommand{\ee}{\end{equation}}
\newcommand{\ba}{\begin{array}}
\newcommand{\ea}{\end{array}}
\newcommand{\bea}{\begin{eqnarray}}
\newcommand{\eea}{\end{eqnarray}}

\newcommand{\it}{\itshape}
\title{Scatophobic Dark Matter}
\author{Gerald Xavier
Gilbert-Thorple$^{1,2}$\footnote{gxgilthor@gmail.com}, 
J\^os\'e Jesus Jesus$^3$\\ 
  \textit{$^1$Kavli University of Southern North Dakota,
Hoople, ND, USA}\\
   \textit{$^2$CERN, Theoretical Physics Department, Geneva, Switzerland}\\
  \textit{$^3$Beijing University, Athens Campus, 
Athens, Greece}
}
\date{}

\begin{document}
%\rightline{CERN-TH-2024-016}
\maketitle

\begin{abstract}
An outstanding mystery of dark matter physics is the lack of direct
detection signals to date.  We suggest that dark matter is {\it
scatophobic}: due to a repulsive long-range interaction, it is
repelled by objects with a large net {\it scat} charge, such as the
Earth, and is therefore not able to reach direct detection
experiments.  This represents the first step in a broader theoretical
paradigm that we dub the ``anti-anthropic principle.''
\end{abstract}

The existence of dark matter in the universe is widely acknowledged,
but it has been questioned whether particle dark matter is the correct
paradigm \cite{Mannheim:2005bfa,Sivaram:2020xih,Nesti:2023tid,
Misiaszek:2023sxe,Schombert:2021xqm,Sanders:2019blc,Melia:2022itm,Gupta:2024eqo,
Milgrom:2009an,Frampton:2023cjz} \cite{Khlopov:2022hjm,Khlopov:2019ibz,Khlopov:2019gnn,Khlopov:2019qcr,
Khlopov:2018ttr,Khlopov:2017bne,Khlopov:2017vcj,Khlopov:2015nrq,Khlopov:2014nva,
Khlopov:2014bia,Khlopov:2013ava,Khlopov:2011tn,Khlopov:2011uy,Khlopov:2011me,
Khlopov:2011zz,Khlopov:2010cr,Khlopov:2010ik,Khlopov:2010jn,Khlopov:2010pq,Khlopov:2009gd,
Khlopov:2009hi,Khlopov:2008zza,Khlopov:2008ty,Khlopov:2008ki,Khlopov:2008rp,Khlopov:2008rq,
Khlopov:2007zz,Khlopov:2006uv,Khlopov:2005ew,Khlopov:1998uj}
\cite{Novikov:2016hrc,Novikov:2016fzd,Novikov:2007xt,Novikov:2005nv,
Novikov:2005gz,Novikov:1969uch}
due to the lack of evidence from
decades of direct searches.  One possibility is to eliminate dark
matter by means of modifying gravity \cite{Zhao:2006vm}, but this comes with a
price: the need for including dark matter \cite{Starkman:2011gpu}.  An amusing suggestion, but
hardly credible, is that dark matter has only gravitational-strength
interactions \cite{Aoki:2024jhr,Barman:2022qgt,Hertzberg:2019bvt,
Garny:2018grs,Ema:2018ucl}.  Alternatively, it may be argued that in fact dark matter
{\it has} been directly detected, with a statistical significance of 60$\,\sigma$
\cite{DAMA:2010gpn}; however it is possible to explain this signal with different
 new physics hypotheses, not requiring dark matter \cite{Gilbert-Thorple:2013ppa}.

The lack of direct detection is a serious criticism, which deserves a
serious answer.  In this {\it Letter}, we propose a dynamical
mechanism that can explain this puzzle, even in the presence of
significant couplings between dark matter (DM) and standard model
particles.  Namely, the DM experiences scatophobic\footnote{The name
is inspired by the well-known model of Ref.\ \cite{Farzan:2012sa}} interactions that
repel it from the vicinity of Earth, before it is able to interact
with direct search experiments.  We will show that this can be
achieved if there is a sufficiently light mediator between the DM
and the repulsive material in the Earth, with reasonable values of the
couplings.  

\begin{figure}[t]
\centerline{\includegraphics[width=0.4\columnwidth]{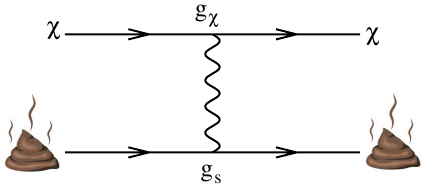}}
\caption{Feynman diagram for dark matter-scat interaction, in the case of a vector mediator.
For a scalar mediator, the wavy line would be replaced by a dashed line.}
\label{fig:diag}
\end{figure}

There are two simple possibilities for such a repulsive interaction, mediated either by
a light scalar or a vector of mass $\mu$ (see Fig.\ \ref{fig:diag}).  For simplicity we will model the scat distribution
on Earth as being spherically symmetric, confined to the surface.  Then dark matter particles
$\chi$ outside of the Earth radius $r_\oplus$ will experience a Yukawa potential given by
\be
	V_s = N_s g_s g_\chi {e^{-\mu r}\over r}
\label{Vseq}
\ee
where $g_s$ and $g_\chi$ are the couplings of the mediator to scat and $\chi$ particles,
respectively, and $N_s$ is the number of scat particles on the Earth's surface.  The sign of 
$g_s g_\chi$ is chosen so that $V_s>0$: the interaction is repulsive.  In the case of a
vector mediator, this requires the dark matter to be asymmetric.  Clearly, DM traveling with
a speed $v_\chi$ will only be able to reach the Earth's surface if $\sfrac12 m_\chi v_\chi^2 >
V(r_\oplus)$.  

If the intrinsic cross section for DM-nucleon interactions is $\sigma_0$, then the effective
cross section relevant for direct detection is given by
$\sigma_{\hbox{\scriptsize eff}} = f_\chi \sigma_0$, where $f_\chi$ is the fraction of DM particles with
energy great enough to overcome the barrier $V(r_\oplus)$.  Let us assume that $\sigma_0$ is
large enough to satisfy the DM naysayers, {\it e.g.,} $\sigma_0 = 1\,$b $=
10^{-24}\,${cm}$^2$.  This may sound very small to a DM naysayer, but they will have to take
our word for it, it is actually very big. Then to avoid the strongest direct detection bounds
(see for example Refs.\ \cite{XENON:2023cxc,LZ:2022lsv}), we should take $f_\chi\lesssim 10^{-23}$.
Assuming a Maxwellian DM velocity distribution with dispersion $v_0\sim 170\,$km/s
\cite{Jiao:2023aci,Brown:2009nh}, and a
cutoff velocity $v_c$,
\be
	f_\chi = \sqrt{2\over \pi}\int_{v_c/v_0}^\infty x^2 e^{-x^2/2}\,dx\,,
\ee
one finds that $v_c \sim 10.5\,v_0\sim 0.06\,c$.  This estimate is likely to be too high
since it exceeds the escape velocity of the Milky Way, $v_{\hbox{\scriptsize esc}}\cong570\,$km/s;  the true cutoff
velocity is expected to be somewhere between $v_{\hbox{\scriptsize esc}}$ and $v_c$.

For simplicity, we initially assume that the mediator mass $\mu\ll 1/r_\oplus \cong 3\times 10^{-14}$\,eV;
then the exponential factor in Eq.\ (\ref{Vseq}) can be ignored.  Taking the
DM mass to be $\sim 30$\,GeV, we find that
\be
	N_s g_s g_\chi \gtrsim \sfrac12 m_\chi v_c^2 r_\oplus \cong 6\times 10^{19}
\label{couplings}
\ee
in order to sufficiently repel the DM particles away from Earth.

The next step is to determine $N_s$, the number of scat particles.  It is determined by the
Master Scat Equation (MSE),
\be
	\dot N_s = N_h \dot n_s - \bar\Gamma_s N_s
\ee
where $N_h$ is the number of humans,\footnote{We ignore the small contributions from other
species \cite{bsfitter}} $\dot n_s$ is the average rate of scat production per person, and
$\bar\Gamma_s$ is the average scat decay rate, including natural and induced (at treatment
plants) decays.  To facilitate future predictions, one should also include the evolution 
equation for $N_h$ and solve the coupled equations for $N_h(t)$.  We take
$N_h(t) = 4.2\times 10^{11}(t_c-t)^{-1.045}$ \cite{pop} where $t$ is time measured in years, 
$t_c = 2062.7$\,y, $\dot n_s = 6\times 10^{24}$/d,\footnote{This corresponds to 124 g/d, translated to number
of carbon atoms} and $\bar\Gamma = 0.1/$y, considering the weighted proportions of naturally
decaying and stimulated decaying scat populations \cite{sdecay}.  In the steady-state
approximation, $\dot N_s = 0$, we obtain
\be
	N_s = {N_h\over \bar\Gamma_s} \dot n_s = 1.1\times 10^{33}
\label{NS0}
\ee
for the present value of $N_s$.  The natural expectation that $g_s = g_\chi$ then implies,
from Eq.\ (\ref{couplings}), the lower bound
\be
	g_s = g_\chi \gtrsim 2.3\times 10^{-7}\,,
\label{glimit}
\ee
a quite modest requirement.  

From the result (\ref{glimit}), we see that there is room to relax the simplifying assumption
of a nearly massless mediator.  For large couplings $g_s=g_\chi=1$, the mediator could be as
heavy as $10^{-12}$\,eV.  For dark photons, this is a very interesting mass region, which
could be compatible with explaining the EDGES 21 cm anomaly \cite{Caputo:2020avy} if there is kinetic
mixing of order $\epsilon\sim 10^{-10}$.

It should be noted that the new interactions proposed are compatible with
observational constraints such as the lack of dark matter scattering in the
Bullet Cluster \cite{Clowe:2006xq}.   These bounds allow for a subdominant component
of dark matter to be strongly interacting, or alternatively for all of the dark matter
to interact strongly with a subdominant component of the baryons.  It is not expected
that more than 10\% of galactic cluster matter will be in the form of scat. 

\begin{figure}[t]
\centerline{\includegraphics[width=0.4\columnwidth]{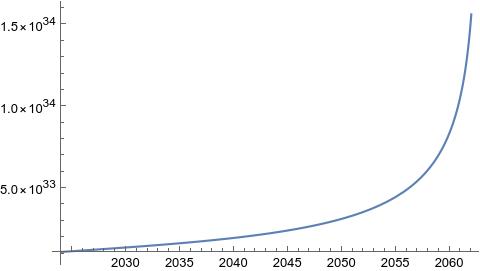}}
\caption{Evolution of the scat number $N_s$ with time.}
\label{fig:Ns}
\end{figure}

To see what the future may hold, we integrate the MSE forward in time
with the initial value (\ref{NS0}), using Mathematica.  The resulting
solution is shown in Fig.\ \ref{fig:Ns}.  It displays a sharp rise in
the coming decades, diverging at a Putin pole near 2063.  This signals
a breakdown of our effective description at late times, which should be
regularized by nuclear war or some other catastrophe.  In any case, it
shows that our result is robust for the near future, guaranteeing that
no direct detection signals will be observed before the demise of
civilization.

The authors of Ref.\ \cite{Nesti:2023tid} argued for a new paradigm, which calls for
more radical developments than simple model-building.  We answer this by proposing
that scatophobic dark matter is just the first manifestation of a broader framework,
that we call the ``anti-anthropic principle.''\footnote{A related
approach was proposed in footnote 7 of Ref.\ \cite{Scott:2006pe}}  Recall that the anthropic principle 
supposes that physical laws must be such as to engender a universe that is compatible
with observers who can perceive and discover such laws.  Here we suppose the opposite:
the Universe should be constructed in such a way as to be hostile toward life,
at least in the form of physicists studying the dark sector.  We believe this will
appeal to the numerous critics of the anthropic principle.  Moreover, it has enormous
potential for explaining null results in a variety of other dark sector searches:
axions, alps, millicharged particles, heavy neutral leptons \dots  We are actively
pursuing the consequences of our new principle in these areas.

\section*{Acknowledgment}
We thank authors who requested citations following an earlier version of this paper.
Figure 1 and all of the writing was made with the help of ChatGPT.  GXGT thanks 
Professor Peter Schickele for valuable discussions.
JJJ is supported by the Sino-Greek Research Foundation Research Pittance Award Program,
grant number SGRFRPAP-0.0000531.  GXGT is supported by the NSF Late Researchers
Wind-Down Accelerator Program, grant number LRWDAP-10498576.  He thanks the CERN Theory
Department for their kind hospitality, and the organizers of Moriond for a congenial
atmosphere that inspired this work.

\section*{Conflict of Interest} GXGT declares that the Alfa Romeo 8C2900, used only
for commuting to and from work, was purchased with different funds than those used for the
present research.

\end{document}